\documentstyle[preprint,aps,epsf]{revtex}
\begin{document}
\draft
\title{\bf{Double-Well Potential : The WKB Approximation with Phase Loss
           and Anharmonicity Effect}}
\author{Chang  Soo  Park and Myung Geun Jeong}
\address{Department of Physics,  Dankook University,  Cheonan 330-714,
         Korea}
\author{Sahng-Kyoon Yoo}
\address{Department of Physics, Seonam University, Namwon, Chunbuk 590-711,
         Korea}
\author{D.K. Park}
\address{Department of  Physics, Kyungnam University,
Masan 631-701, Korea}
\date{\today}
\maketitle
\begin{abstract}
We derive a general WKB energy splitting formula in a double-well
potential by incorporating both phase loss and anharmonicity
effect in the usual WKB approximation. A bare application of the
phase loss approach to the usual WKB method gives better results
only for large separation between two potential minima. In the
range of substantial tunneling, however, the phase loss approach
with anharmonicity effect considered leads to a great improvement
on the accuracy of the WKB approximation.
\end{abstract}

\narrowtext

\section{INTRODUCTION}
\label{sec:intro}

The quantum tunneling in the double-well potential, $V(x) =\lambda x^4 -
k x^2 (k, \lambda >0)$, is a longstanding and well-known problem in quantum
mechanics. One of the main interests in this problem is how accurately one
can calculate the energy splitting of degenerate energy levels due to quantum
tunneling. The nonperturbative nature of the double-well potential, which
stems from the quartic term, does not allow standard perturbation treatment.
Instead several alternative methods have been proposed. Among those there
are the instanton method \cite{Kleinert,Gildener}, the WKB approximation
\cite{Landau,Banerjee}, and a numerical method \cite{Banerjee}.

In the instanton method, one uses imaginary time path integral to obtain
the classical action - the so-called euclidean action. Qualitatively, this
method is useful in understanding the quantum tunneling which has no
classical correspondence. However, the quantitative calculation of the energy
splitting in the double-well potential by this method is inaccurate because
the euclidean propagator can be
obtained only in the limit of infinite separation between the two potential
minima \cite{Kleinert}, which corresponds to zero tunneling probability. Thus,
the validity of the instanton approach is restricted to the case of which the
two potential minima are far apart, and its accuracy is expected to be reduced
as they become close.

In the case of WKB method, one uses the semiclassical approximation in which
the de Broglie wavelength $\lambda (x) = h/\sqrt{2m(E-V(x))}$ is assumed to be
very short and varies sufficiently slowly. The main defect of this approximation
comes from the inevitable difficulties in treating the divergence of $\lambda(x)$
at a classical turning point. Recently, a phase loss approach has been proposed
to improve this shortcoming for general barrier potentials \cite{Trost}. In the
usual WKB approximation, based on the short wavelength condition, the phase
change of the WKB wave function at a classical turning point is taken to be
$\pi/2$. However, since the wavelength should not be arbitrarily short, the phase
change should become a nonintegral multiple of $\pi/2$. Based on this idea the
authors of Ref.5 have shown that the accuracy of the WKB wave function is greatly improved.

In this paper we use this phase loss approach to calculate the WKB energy
splitting in the double-well potential. Although the authors of
Ref.5 applied this approach to several barrier potentials, the
double-well potential, which is one of the prototypes in
tunneling problem, was not treated. In addition to this we also consider
the anharmonicity effect of the double-well potential. In most previous works
\cite{Gildener,Banerjee,Bhatta} the energy eigenvalue of a unperturbed
harmonic oscillator has been used for the calculation of the energy
splitting, which corresponds to the case of the infinite separation of the
two potential minima, and hence exposes same defect as the instanton method.
The present work, however, incorporates anharmonicity into the WKB formalism,
which gives a more realistic model, and consequently a more improved result
in the region over which the tunneling is appreciable. By using the phase loss approach
with the anharmonicity effect we will present a general WKB energy splitting formula in a
double-well potential. From this, the usual WKB energy splitting
formula can be obtained by both taking short wavelength limit and using energy
level corresponding to the infinite separation of the potential minima. Also,
if we take the short wavelength limit only we can derive an anharmonic WKB
energy splitting formula. We then present comparisons between numerical results
obtained from these formula ; for completeness, we also include results from
the instanton method. We find that a bare application of the
phase loss approach to the usual WKB method leads to better
results only in the limits of large separation. If we consider the
anharmonicity effect, however, the phase loss approach greatly improves
the accuracy of the usual WKB approximation, especially in
the region where the tunneling is appreciable. In the following section, we will first derive
a general WKB energy splitting formula. In section III, we will show comparisons
between results of various cases and give discussions. Finally, there will
be a summary in section IV.

\section{Energy splitting formula}
\label{Energy}

We consider a double-well potential of the type given in the introduction,
which is shown in Fig.1. The Hamiltonian for this potential is
\begin{equation}
H=p^2 + \lambda x^4 - kx^2
\end{equation}
in which $p$ is the particle momentum, and we have taken the particle mass
$m=1/2$. We will also take $\hbar = 1$ throughout this paper. In the WKB
approximation we have phase integrals in the two classically allowed regions:
\begin{equation}
a= \int_{\alpha}^{\gamma} p_n (x) dx, \hspace{10mm}
b= \int_{-\alpha}^{-\gamma} p_n (x) dx ,
\end{equation}
and inside the central barrier we also have
\begin{equation}
S= \int_{-\alpha}^{\alpha} \kappa_n (x) dx ,
\end{equation}
where
\begin{equation}
p_n (x) = \sqrt{E_n - V(x) } = i \kappa_n (x)
\end{equation}
with $E_n  $ being an energy level in one of the wells. When we include the
anharmonicity effect this energy level  becomes, to second order correction,
\begin{equation}
E_n = E_n^{(0)} + \delta_n ,
\end{equation}
where
\begin{equation}
E_n^{(0)} = (2n + 1 ) \sqrt{2k} - \frac{k^2}{4\lambda}
\end{equation}
is the energy level of a  unperturbed harmonic oscillator and $\delta_n $ is
the second order correction term given as
\begin{equation}
\delta_n = - \frac{3\lambda}{k} (n^2 + n + \frac{1}{3} ) -
\frac{\lambda^2 }{64k^2 \sqrt{2k}} (34n^3 + 51n^2 + 59n + 21 ) .
\end{equation}
To obtain the WKB energy
splitting formula in the double-well potential we follow the same method as
described in Ref.7 except using arbitrary phase losses $\phi_{\alpha}
(= \phi_{-\alpha} ), \phi_\gamma (=\phi_{-\gamma} )$ at each turning point
instead of $\pi/2$ which corresponds to the short wavelength limit. After
long but straightforward calculations we obtain
\begin{equation}
\Delta E_{\rm anp} = \frac{1}{T} \frac{4e^S \tan^2 \frac{1}{2}(\phi_\alpha
+ \phi_\gamma) + 4 e^S }
{4 e^{2S} \tan^2 \frac{1}{2}(\phi_\alpha + \phi_\gamma) - 1} ,
\end{equation}
where
\begin{equation}
T = \left( \frac{ \partial a}{\partial E_n } \right) =
\frac{1}{2} \int_\alpha^\gamma \frac{dx}{p_n (x) } .
\end{equation}
To find the phase losses at the turning points we modify the matching
condition suggested in Ref.5. Since each well is asymmetric the phase changes
at $\alpha$ and $\gamma$ are different. At $x = \pm \gamma$ the phase losses
are given by
\begin{equation}
\phi_\gamma = -2 \tan^{-1} \left[ \frac{1} {p_{\rm{min}}}
\frac{\psi'(\gamma)}{\psi(\gamma)} \right],
\end{equation}
where
\begin{equation}
p_{\rm{min}}  = \sqrt{E_n - V(\pm \sqrt{k/2\lambda})},
\end{equation}
and
\begin{eqnarray}
\psi (\gamma ) &=& \int_{\gamma }^{\infty} (x - \gamma ) \kappa_n^{3/2} (x)
e^{-\int_{\gamma }^x \kappa_n (x')dx' }dx \\
\psi' (\gamma ) &=& - \int_{\gamma }^\infty \kappa_n^{3/2} (x)
e^{-\int_{\gamma }^{x} \kappa_n (x') dx' }dx .
\end{eqnarray}
The phase losses $\phi_{-\alpha}(=\phi_\alpha )$ at $\pm \alpha $ can be
obtained from the following algebraic equation :
\begin{equation}
\frac{\psi' (-\alpha)}{\psi (-\alpha)} = - p_{\rm{min}} \tan \left( a -
\frac{\phi_\gamma }{2} \right) .
\end{equation}
In this case the Lippmann-Schwinger equations should be written by
\begin{eqnarray}
\psi (-\alpha )  &=& \int_{-\alpha}^{\alpha} (x + \alpha) \kappa_n^{3/2}(x)
\psi_1 (x) dx + \int_{\alpha}^{\gamma} (x + \alpha ) p_n^{3/2} (x) \psi_2 (x)
dx \nonumber \\
&+& \int_{\gamma}^{\infty} (x + \alpha ) \kappa_n^{3/2} (x) \psi_3 (x) dx \\
\psi'(-\alpha) &=& \int_{-\alpha}^{\alpha} \kappa_n^{3/2} (x) \psi_1 (x) dx
+ \int_{\alpha}^{\gamma} p_n^{3/2} \psi_2 (x) dx + \int_{\gamma}^{\infty}
\kappa_n^{3/2} (x) \psi_3 (x) dx ,
\end{eqnarray}
where $\psi_1 (x) , \psi_2 (x), \psi_3 (x) $ are the WKB wavefunctions inside
the central barrier, in the right well, and in the right-barrier, respectively.
By applying the WKB connection formula at each turning point we can obtain
\begin{eqnarray}
\psi_1 (x) &=& \cos A ~ e^{-\int_{-\alpha}^{x} \kappa_n (x')dx' } - 2 \sin A ~
e^{\int_{-\alpha}^{x} \kappa_n (x') dx' } \\
\psi_2 (x) &=& e^{-S} \cos A~ \sin \left( \frac{\phi_\alpha}{2} -
\int_{\alpha}^{x} p_n (x')dx'\right) - 4e^S \sin A~ \cos \left(
\frac{\phi_\alpha}{2} -
\int_{\alpha}^{x} p_n (x')dx' \right) \\
\psi_3 (x) &=& - \left[ \frac{1}{4} e^{-S} + e^S \right] \sin 2A ~
e^{-\int_{\gamma}^{x} \kappa_n (x')dx' }
\end{eqnarray}
with
\begin{displaymath}
A = a - \frac{\phi_\alpha + \phi_\gamma }{2}.
\end{displaymath}
We note that Eq.(8) is a general WKB energy splitting formula with both
anharmonicity effect and phase loss.
In this formula, if we use the unperturbed energy level $E_n^{(0)}$ we have
the WKB formula with phase loss only:
\begin{equation}
\Delta E_{\rm up} = \frac{1}{T^{(0)}} \frac{4e^{S^{(0)}} \tan^2 \frac{1}{2}
(\phi_\alpha + \phi_\gamma) + 4 e^{S^{(0)}} }
{4 e^{2S^{(0)}} \tan^2 \frac{1}{2}(\phi_\alpha + \phi_\gamma) - 1} ,
\end{equation}
where
\begin{displaymath}
T^{(0)} = \frac{1}{2} \int_{\alpha}^{\gamma} \frac{dx}{p_n^{(0)} (x)} ,
\hspace{10mm} S^{(0)} = \int_{-\alpha}^{\alpha} \kappa_n^{(0)} (x) dx
\end{displaymath}
with
\begin{displaymath}
p_n^{(0)} = \sqrt{E_n^{(0)} - V(x)}.
\end{displaymath}
Also, If we let $\phi_\alpha = \phi_\gamma = \pi/2 $ in Eq.(8), it reduces to
\begin{equation}
\Delta E_{\rm an} = \frac{1}{T} e^{-S} ,
\end{equation}
which corresponds to the case of the WKB approximation with anharmonicity
effect only. Finally, by both taking $\phi_\alpha = \phi_\gamma = \pi/2 $
and using Eq.(20) the usual WKB energy splitting formula can be obtained to be
\begin{equation}
\Delta E_{\rm u} = \frac{1}{T^{(0)}} e^{-S^{(0)}} .
\end{equation}

\section{NUMERICAL RESULTS AND COMPARISONS}
\label{RESULTS}

To examine whether and how much the general WKB energy splitting formula
improves the WKB approximation we calculate the various energy splittings
derived in the previous section numerically and compare each numerical result
with the exact values obtained by the nonperturbative method [4].  For
completeness, we also include numerical results from the instanton method by
which the energy splitting is obtained to be [2]
\begin{equation}
\Delta E_{\rm in} = \left[ \frac{64 k^2 \sqrt{2k}}{\pi \lambda} \right]^
{\frac{1}{2}} e^{- \frac{k \sqrt{2k}}{3\lambda}} .
\end{equation}
The numerical calculations have been performed for $n = 0$ (the ground state)
case with $k=1$, and we find $\Delta E$'s for different $\lambda$'s ; the $n=1$
case does not give enough values for comparisons. For
direct comparisons, we compute the ratios of each numerical result and the
exact value taken from Ref.4, then find the respective differences of the
ratios from 1. The final results are listed in Table 1.

An immediate observation from this table is, as remarked in the introduction,
that both the instanton (the fifth column) and the usual WKB (the first
column) methods give good estimate only for small values of $\lambda$ which
correspond to the limiting case of infinite separation of the two potential
minima, and their results are getting worse as $\lambda$ increases \cite{abo}.
It is also clear that the usual WKB approximation is better than the
instanton method, which is consistent with the result of previous work [4].

For the presence of tunneling in the double-well potential should the
separation between the two potential minima be neither too small nor infinitely
large. When the two minima are too close the height of the barrier  becomes
lower than the lowest energy level in each well. This can be seen from the
first four columns where energy splittings for $\lambda = 0.2$ do not appear.
In the opposite limit of $\lambda \rightarrow 0$ the two potential wells
become two independent harmonic oscillator potentials, which does not allow
tunneling. In order for the tunneling probability not to vanish the
two potential wells should not be infinitely apart, which requires the
consideration of the anharmonicity effect in calculating the energy splittings.
From Eq.(5) and Eq.(7) this effect is seen to be little for small values of
$\lambda$. The calculation of the energy splitting due to tunneling in this
region is, then, governed by the usual WKB formula. As $\lambda$ increases,
however, the two wells approach each other, and the anharmonicity effect
become more important. Especially, for an appreciable tunneling probabiliy
the anharmonic property of the double-well potential should be incorporated
in the calculation of the energy splitting. Thus, the result from the WKB
formula with anharmonicity effect is expected to be improved as $\lambda$
increases. We can see this behavior from the comparisons of the first column
and the third one (the WKB result with anharmonicity only). For $\lambda
\leq 0.05$ the usual WKB results are more accurate than that with
anharmonicity effect. But, for $\lambda \geq 0.07$ the results from
the WKB formula  with anharmonicity effect are much more accurate than
the usual WKB result. Note that for $\lambda = 0.15$ the error of the
anharmonic result is less than $2\%$, while that of the usual WKB result
has almost $20\%$. This illuminates that the usual WKB approximation
becomes unreliable for small separation distance.

We now compare the first column with the second one. For small values of $
\lambda (\leq  0.03 )$, the phase loss approach is better than the usual WKB
estimate. For $\lambda \geq 0.04$, however, the result from the phase loss is
more inaccurate than the usual WKB result. By comparing the third column with
the forth one we also find that the results from the general energy splitting
formula which includes both the phase loss and the anharmonicity effect is
slightly better than those from the WKB method with anharmonicity effect only.
According to Ref.5 the calculation of phase loss at turning point is based on the
assumption that the WKB wavefunction inside a well is proportional to the
exact wavefunction which can be obtained from Schr\"{o}dinger equation
around the potential minimum. The condition for the validity of this
approximation is that the dimensionless parameter $p_{\rm{min}} x_0 $, where
$x_0 $ being the distance between the point of potential minimum and the
turning point, should be small. In the case of usual WKB method, while $x_0 $
increases with $\lambda$, $p_{\rm{min}} = \sqrt{E_0^{(0)}}$ is independent of
$\lambda$ so that $p_{\rm{min}} x_0 $ becomes larger with increasing $\lambda$.
Thus, the application of the phase loss approach to this case is expected to be
valid only for small $\lambda's $. On the other hand, when the anharmonicity
effect is included, from Eq.(11) $p_{\rm{min}}$
decreases with $\lambda$. Although the $x_0 $ value in anharmonic case also
increases with $\lambda$ the dimensionless
parameter $p_{\rm{min}} x_0 $ remains small even for large values of $\lambda$.
Consequently, the incorporation of the anharmonicity effect in the
WKB approximation makes the phase loss approach valid even for large values of
$\lambda$. Therefore, the entries in the second and the forth columns are reasonable
results.

Finally, we discuss the occurence of a larger energy splitting in phase loss
approach than in the usual (or anharmonic) WKB approximation. If we compare
the first column (or the third column) with the second one (or the forth one)
we can notice that the energy splitting $\Delta E_{\rm anp}$ (or
$\Delta E_{\rm up}$) obtained by the phase loss approach is larger than
$\Delta E_{\rm an}$ (or $\Delta E_{\rm u}$).
The reason for this can be explained as follows. Since
the usual WKB approximation is based on the condition of short wavelength
limit its formalsim is more particlelike. Thus, in this semiclassical limit,
the tunneling effect, which originates from pure wave property, is ill
treated. However, in the phase loss approach we use the WKB method away from
the short wavelength limit, and hence the approach is more wavelike.
Therefore, the phase loss approach treats the tunneling effect better, by
which a larger energy splitting follows.

\section{SUMMARY}
\label{summary}
We have investigated the phase loss approach in the WKB energy
splitting formula about a double-well potential. By incorporating
the phase loss approach with the anharmonicity effect in the
usual WKB approximation we have obtained a general WKB energy
splitting formula. A bare application of the phase loss to the
usual WKB approximation yields a slight improvement only in the
limits of large separation of the two potential minima. When the
anharmonicity effect is included, however, the phase loss approach
significantly improves the accuracy of the usual WKB method, especially
in the range of appreciable tunneling.

In applying the phase loss approach to the WKB approximation it is
crucial to calculate the phase change at turning point accurately,
for which there exists a condition such that the dimensionless
parameter $p_{\rm{min}} x_0$ is small. As noted in Sec.III the
inclusion of the anharmonicity  keeps this condition over a
region where otherwise the condition becomes invalid,
so that a correct phase change at turning point can be
obtained. It is thus important to consider the anharmonicity
effect in the application of the phase loss approach to an
anharmonic potential such as double-well potential.

Finally, We have also found that the
magnitude of energy splitting obtained from phase loss approach is
larger than that of the WKB energy splitting obtained in the limit
of short wavelength.

\begin{figure}
\caption{ One-dimensional double-well potential. $E$ is the ground state
enrgy, and $\pm \alpha$, $\pm \gamma$ are the classical turning points
corresponding to $E$.}
\end{figure}

\newpage
\epsfysize=20cm \epsfbox{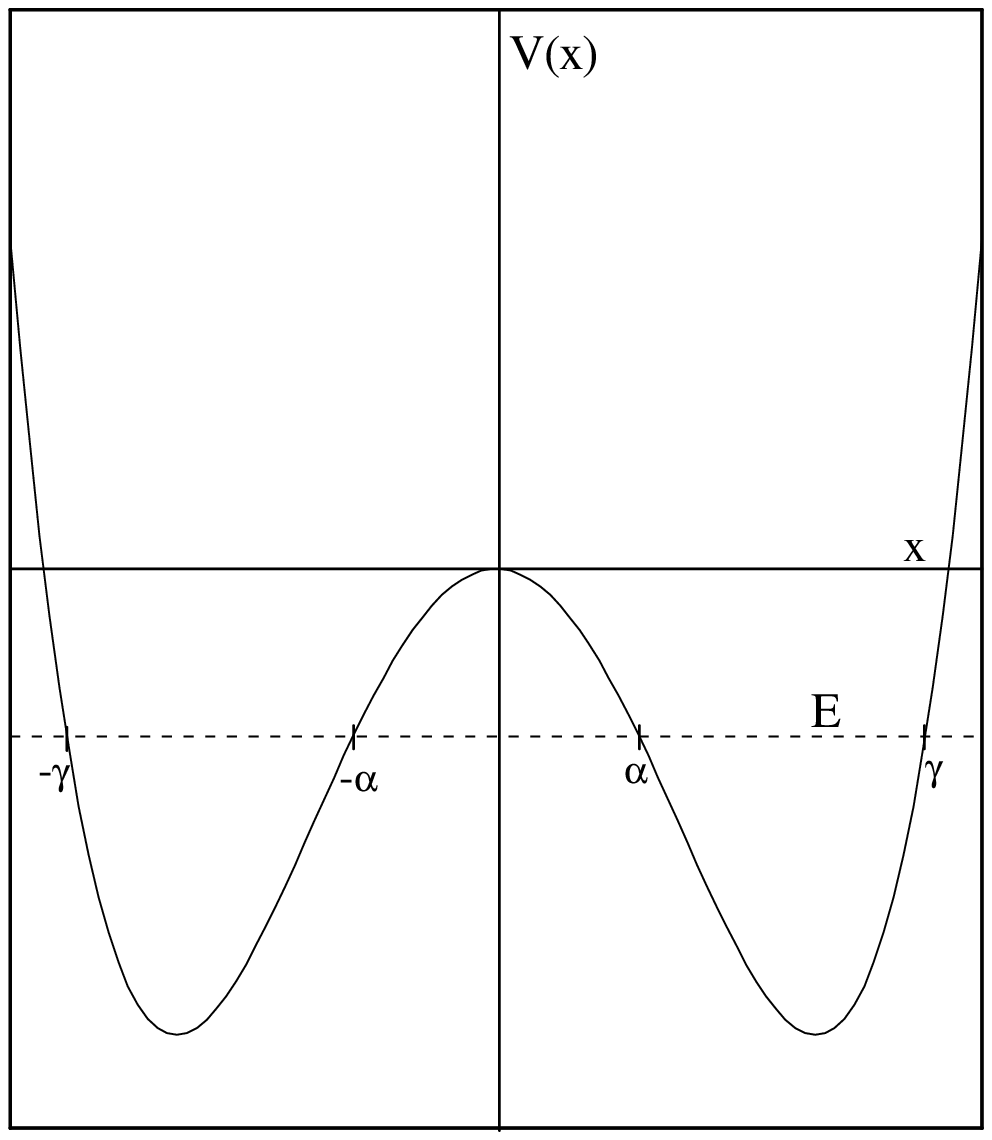}
\begin{table}
\centering
\caption{Ratios of numerical result and the exact value taken from Ref.4.  All
entries are the differences of the ratios from 1.}
\vspace{1cm}
\begin{tabular}{cccccc}
{ $\lambda$ }
& {$\frac{\Delta E_{\rm{u}}}{\Delta E_{\rm{exact}}} - 1$}
& {$\frac{\Delta E_{\rm{up}}}{\Delta E_{\rm{exact}}} - 1$}
& {$\frac{\Delta E_{\rm{an}}}{\Delta E_{\rm{exact}}} - 1$}
& {$\frac{\Delta E_{\rm{anp}}}{\Delta E_{\rm{exact}}} - 1$}
&{ $\frac{\Delta E_{\rm{in}}}{\Delta E_{\rm{exact}}} - 1$ }\\
\hline
0.02  & -0.03075 & -0.03046 & -0.07359 & -0.07357 & 0.04524\\
0.03  & -0.01297 & -0.01227 & -0.07382 & -0.07376 & 0.07020\\
0.04  & 0.00543 & 0.00676 & -0.07263 & -0.07252 & 0.09773\\
0.05  & 0.02466 & 0.02691 & -0.07006 & -0.06990 & 0.12809\\
0.07  & 0.07348 & 0.07883 & -0.05426 & -0.05396 & 0.20694\\
0.1   & 0.15125 & 0.16762 & -0.02278 & -0.02157 & 0.35392\\
0.15  & 0.21684 & 0.25018 & 0.01799 & 0.01829 & 0.66831\\
0.17  & 0.07471 & 0.08135 & -0.00882 & -0.00804 & 0.77935\\
0.2   & none    & none    & none    & none    & 0.91421
\end{tabular}
\end{table}


\begin{references}
\bibitem{Kleinert} Hagen  Kleinert, {\it{PATH  INTEGRALS in  Quantum Mechanics,   Statistics,
and Polymer Physics}} (World Scientific, Singapore, 1990), Chap. 17.

\bibitem{Gildener} Eldad Gildener and Adrian Patrascioiu, Phys.  Rev. D
{\bf{16}}, 423 (1977).

\bibitem{Landau} L. D. Landau and E. M. Liftshitz, {\it{Quantum Mechanics}},
3rd ed. (Pergamon, New York, 1977), Chap. VII.

\bibitem{Banerjee} K. Banerjee and P. Bhatnagar, Phys. Rev. D {\bf{18}},
4767 (1978).

\bibitem{Trost} H. Friedrich and J. Trost, Phys. Rev. Lett. {\bf{76}},
4869 (1996); H. Friedrich and J. Trost, Phys. Rev. A {\bf{54}}, 1136 (1996).

\bibitem{Bhatta} S. K. Bhattacharya, Phys. Rev. A {\bf{31}}, 1991 (1985);
S. K. Bhattacharya  and  A. R. P. Rau, Phys. Rev. A {\bf{26}},  2315 (1982).

\bibitem{Park} David Park, {\it{Introduction to the Quantum Theory}}, 3rd ed.
(McGraw-Hill, Inc. 1992), Chap. 4, p. 115.

\bibitem{abo} In the case of the usual WKB approximation, since the
antisymmetric energy level for $\lambda = 0.17$ lies above the barrier
maximum we have unexpected result which is better than the result for
$\lambda = 0.15$.


\end{references}
\end{document}